\documentstyle[prd,aps]{revtex}
\begin{document}
\input epsf
\draft
\twocolumn[\hsize\textwidth\columnwidth\hsize\csname
@twocolumnfalse\endcsname
\preprint{SU-ITP-98-45, hep-ph/9807493}
\title{A Toy Model for Open Inflation}
\author{Andrei Linde}
\address{Department of Physics, Stanford University, Stanford, CA
94305, USA}
\date{July 24, 1998}
\maketitle
\begin{abstract}
The open inflation scenario based on the theory of  bubble formation   in the models of a single scalar field suffered from a fatal defect. In all the versions of this scenario known so far, the Coleman-De Luccia instantons describing the creation of an open universe did not exist. We propose a simple one-field model where the CDL instanton does exist and the  open inflation scenario can be realized.  
\end{abstract}
\pacs{PACS: 98.80.Cq  \hskip 3.6cm SU-ITP-98-45
\hskip 3.6cm  hep-ph/9807493}
\vskip2pc]

\section {Introduction}

Until very recently, we did not have any consistent
cosmological models describing a homogeneous open universe. Even though the
Friedmann open universe model did exist, it  did not appear to
make any sense to assume that all parts of an infinite universe can be
created simultaneously  and have the same value of energy density everywhere.

This problem was solved only after the invention of inflationary cosmology.  (This is somewhat paradoxical, because most of inflationary models predict that the universe must be flat.) The main idea was to use the well known fact that the bubbles created in the process of
quantum tunneling tend to have
spherically
symmetric shape, and homogeneous interior, if the tunneling probability is
strongly suppressed.  Bubble formation in the false vacuum is described by the Coleman-De Luccia (CDL) instantons \cite{CL}. Any bubble formed by this mechanism looks from inside like an infinite open
universe \cite{CL,Gott}.
 If this universe continues inflating
inside the bubble, then we obtain an open inflationary universe. Then by a certain fine-tuning of parameters one can get any value of $\Omega$ in the range $0< \Omega < 1$ \cite{BGT}.

Even though the basic idea of this scenario was pretty simple, it was very difficult to find a realistic open inflation model. The general scenario proposed in \cite{BGT}  was based on investigation of chaotic inflation and
tunneling in the theories of one scalar field $\phi$.  There are many papers containing a detailed investigation of density perturbations, anisotropy of the microwave background radiation, and gravitational wave production in this scenario.  However, no models where this scenario could be successfully realized have been proposed so far. 
As it was shown in \cite{Open},
in the simplest models with polynomial potentials of the type of
${m^2\over 2} \phi^2-{\delta\over 3} \phi^3 + { \lambda\over 4} \phi^4$ the
tunneling  occurs not  by bubble
formation, but by jumping onto the top of the potential barrier described by
the
Hawking-Moss instanton \cite{HM}.   This process leads to  formation of inhomogeneous
domains of a
new phase, and the whole scenario fails. As we will show in this paper, the main reason for this failure is rather generic. Typically, CDL instantons exist  only  if $|V''| > H^2$ during the tunneling. Meanwhile, inflation, which, according to \cite{BGT}, begins immediately after the tunneling, typically requires $|V''| \ll H^2$. These two conditions are nearly  incompatible.

This problem can be avoided if one considers models of two
scalar fields \cite{Open}.  In this scenario the bubble formation occurs due to tunneling with respect to one of the fields which has   a steep barrier in its
potential. Meanwhile   the role of the inflaton inside the bubble is played by another field, rolling along
a flat direction ``orthogonal'' to the direction of
quantum tunneling. Inflationary models of this type  have many interesting features. In these models the universe consists of
infinitely many expanding bubbles immersed into an
exponentially expanding false
vacuum state. Interior of each of these bubbles looks like an infinitely large open
universe, but the values of $\Omega$ in these universes may take any value from
$1$ to $0$.

Many versions of these two-field models have been considered in the recent
literature, for a review see
e.g. \cite{Bellido}. Strictly speaking, however, the two-field models describe quasi-open universes rather than the open ones. The reason why the interior of the bubble in the one-field model can be associated with an open universe is based on the possibility to use this field as a clock, which is most suitable for the description of the processes inside the bubble from the point of view of an internal observer. If one has two fields, they are not always perfectly synchronized, which may lead to deviations of the internal geometry from the geometry of an open universe \cite{Open} and may even create  exponentially large quasi-open regions with different $\Omega$ within each bubble \cite{GarrBell}. This makes a complete quantum mechanical investigation of these models rather involved.  

Recently an attempt has been made to describe the quantum creation of an open universe in the one-field models of chaotic inflation with the simplest potentials of the type of $\phi^n$ without any need for the Coleman-De Luccia bubble formation \cite{HT}.  Unfortunately, all existing versions of this scenario lead either to a structureless universe with $\Omega = 10^{-2}$ \cite{HT,ALOpen}, or to the universe where inflation is impossible \cite{Duff}. The only  exception is the model proposed by Barvinsky, which is based on investigation of the one-loop effective action in a theory of a scalar field with an extremely large nonminimal coupling to gravity \cite{Barv}. However, this model, just as the original model of Ref. \cite{HT}, is based on the assumption that the quantum creation of the universe is described by the Hartle-Hawking wave function \cite{HH}. Meanwhile, according to   \cite{ALOpen,Creation,book,LLM}, the Hartle-Hawking wave function describes the ground state of the universe (in those cases when such a state exists) rather than the probability of the quantum creation of the universe. Instead of describing the creation of the universe and its subsequent relaxation to the minimum of the effective potential, which is the essence of inflationary theory, it assumes that a typical universe from the very beginning is in the ground state corresponding to the minimum of $V(\phi)$. This is the main reason why the Hartle-Hawking wave function fails to predict a long stage of inflation and reasonably large $\Omega$ in most of inflationary models. Another problem of this scenario is related to the singular nature of the Hawking-Turok   instanton \cite{ALOpen,VIL}.

Thus, in our opinion, until now we did not have any simple and satisfactory one-field open universe model predicting $0.2 < \Omega < 1$, neither of the type of \cite{BGT}, nor of the type of \cite{HT}. The two-field models \cite{Open} of the desirable type do exist, but they are often rather complicated to analyse, so they should be studied case by case.  

The purpose of this paper is to go back to basics and  re-examine the possibility of the  one-field open inflation due to the CDL tunneling. First of all, we will explain  why it was so difficult to realize this scenario. Then we will give an example of a model where this can be accomplished.  We do not know whether our model is realistic, but it is so simple that for the time being it can serve as a toy model for  open inflation.

\section{Problems with the one-field open inflation}\label{problems}

Suppose we have an effective potential $V(\phi)$ with a local minimum
at $\phi_0$, and a global minimum at $\phi=0$, where $V=0$.  In an $O(4)$-invariant Euclidean spacetime with the
metric
\begin{equation}\label{metric}
ds^2 =d\tau^2 +a^2(\tau)(d \psi^2+ \sin^2 \psi \, d \Omega_2^2) \ ,
\end{equation}
the scalar field $\phi$ and the three-sphere radius $a$ obey the
equations of motion
\begin{equation}\label{eq1}
\phi''+3{a'\over a}\phi'=V_{,\phi} \ ,  ~~~   a''= -{8\pi \over 3} a (
\phi'^2 +V) \ ,
\end{equation}
where primes denote derivatives with respect to $\tau$.
Here and in what follows we will use the units where $M_p = G^{-1/2} = 1$.

An instanton which describes the creation of an open universe
was first found by Coleman and De Luccia   \cite{CL}.  It is
given by a slightly distorted de~Sitter four-sphere of radius
$H^{-1}(\phi_0)$, with $a \approx H^{-1} \sin H \tau$. Typically, the field $\phi$ is very close to the
false vacuum, $\phi_0$, throughout the four-sphere except in a small
region (whose center we may choose to lie at $\tau=0$), in which it
lies on the `true vacuum' side of the maximum of $V$. The scale factor $a(\tau)$ vanishes at the points
$\tau=0$ and $\tau=\tau_{\rm i} \approx \pi/H$. In order to get a
singularity-free solution, one must have $\phi' = 0$ and $a'=\pm 1$ at $\tau=0$ and $\tau=\tau_{\rm i}$. This configuration interpolates between some initial point $\phi_i \approx = \phi_0$ and the final point $\phi_f$ \cite{min}.  After a proper analytic continuation to the Lorentzian regime, it describes an expanding   bubble  which  contains   an   open universe \cite{CL}. 

Solutions of this type can exist only if the bubble can fit into   de Sitter sphere of radius  $H^{-1}(\phi_0)$. To understand whether this can happen, remember that at  small $\tau$ one has $a \sim \tau$, and Eq. (\ref{eq1}) coincides with equation describing creation of a bubble in Minkowski space, with $\tau$ being replaced by the bubble radius $r$: \ $\phi''+ {3 \over r}\phi'=V_{,\phi}$ \cite{Coleman}. Here the radius of the bubble can run from $0$ to $\infty$. Typically the bubbles have size greater than the Compton wavelength of the scalar field, $r \gtrsim m^{-1} \sim (V'')^{-1/2}$ \cite{nucl}.  

In de Sitter space $\tau$ cannot be greater than ${\pi\over H}$, and in fact the main part of the evolution of the field $\phi$ must end at $\tau \sim {\pi \over 2 H}$. Indeed, once the scale factor reaches its maximum at $\tau \sim {\pi \over 2 H}$, the coefficient ${a'\over a}$ in Eq. (\ref{eq1})  becomes negative, which corresponds to anti-friction. Therefore if the field $\phi$  still changes rapidly at $\tau > {\pi \over 2 H}$, it  experiences ever growing acceleration near $\tau_{\rm f}$, and typically the solution becomes singular \cite{HT}.
Thus one may expect that Coleman-De Luccia (CDL) instantons are possible only if ${\pi \over 2 H} >   (V'')^{-1/2}$, i.e. if $H^2 < V''$. It is important that this condition must be satisfied at small $\tau$, which correspond to the final point to which the tunneling should bring us, and where inflation should begin in accordance with the scenario of Ref. \cite{BGT}.  But this condition is  opposite to the standard inflationary condition $H^2 \gg V''$. This suggests that the basic mechanism of \cite{BGT} simply cannot work: Tunneling with bubble production can only bring us to a very curved part of the effective potential, where inflation is impossible.

To be more accurate, the condition $H^2 \gg V''$ can be violated during sufficiently short intervals of the rolling of the field field $\phi$. But then during these intervals the usual inflationary fluctuations of the scalar field $\phi$ are not produced. Thus, even if   inflation begins 
after the CDL tunneling, one should expect suppression of those density perturbations which are produced inside the bubble soon after the tunneling. This effect has not been anticipated in the previous studies of the one-field open inflation.

To illustrate our general arguments, we will consider
 the most natural possibility, which is the  chaotic
inflation in the model with the effective potential \cite{Open}
\begin{equation}\label{o1}
V(\phi) = {m^2\over 2} \phi^2 - {\delta\over 3} \phi^3 + {\lambda\over 4}\phi^4
\ .
\end{equation}
In order to obtain an open inflationary
universe in this model it is necessary to adjust   parameters  so that that the
tunneling creates bubbles with   $\phi \sim 3 M_p$. In such a case the
interior of the bubble after its formation inflates by about $e^{60}$ times,
and $\Omega$ at the present time may become equal to, say, $0.3$. The local minimum of the effective potential in this model appears at $\phi = {\delta \over 2 \lambda} + \gamma$, where
 $ \gamma = \sqrt{  {\delta^2 \over 4 \lambda^2} - {m^2\over \lambda} }$.
The local  minimum of the effective potential appears for ${\delta > 2
\sqrt\lambda\, m }$, and it becomes unacceptably deep (deeper than the minimum
at $\phi = 0$) for  ${\delta > {3 \sqrt\lambda\over \sqrt 2}\, m }$.  Thus in
the whole region of interest one can use a simple estimate ${\delta \sim  2
\sqrt\lambda\, m }$ and represent $\gamma$ in the form $\beta{m\over 2
\sqrt{2\lambda}}$, with  $\beta < {1}$.  The local maximum of the potential
appears at $\phi = {\delta \over 2 \lambda} - \gamma$. Tunneling should occur
to some point with $3   < \phi < {\delta \over 2 \lambda} - \gamma$, which
implies that ${\delta \over  \lambda} > 6  $.

The best way to study tunneling in this theory is to introduce the field $\chi$
in such a way that $\chi = 0$ at the local minimum of $V(\phi)$: $\chi = -\phi +{\delta \over 2 \lambda} + \gamma$. Then one can show that  if the local minimum is not very deep
($\beta \ll 1$), the effective potential (\ref{o1}) can be represented as
\begin{equation}\label{o5}
V(\chi) \approx {m^2\delta^2 \over 48\lambda^2} + { \sqrt 2\beta m^2 } \chi^2 -
{\delta\over 3} \chi^3 + {\lambda\over 4}\chi^4 \ .
\end{equation}
The Hubble constant in the local minimum  is given by
$H^2 \sim  { \pi \delta^2  m^2\over 18\lambda^2  } >  {2\pi}\, m^2$, which
is much greater than the effective mass squared of the field $\chi$ for
$\beta \ll 1$, $V'' = {2\sqrt 2\beta m^2}$. Therefore CDL instantons do not exist in this regime.

A simple way to understand this problem is to note that for any potential $V(\phi) \sim \phi^n$ one has $H^2 \gg V''$ for $\phi > 1$. This relation remains true for a generic polynomial potential unless different terms $\phi^n$  cancel   each other and $V(\phi)$ becomes very small near its local minimum.
Thus, a possible way to overcome this problem would be to consider the case ${\delta
\approx {3 \sqrt\lambda\over \sqrt 2}\, m }$\, ($\beta \approx 1$). Then the
two minima of the effective potential become nearly degenerate in energy density, and
$H^2$ becomes much smaller than $V''$ in the false vacuum.  However, this does not mean that CDL instantons exist even in this degenerate case. Indeed, the curvature of the effective potential in this case will remain much smaller than $H^2$ near the local maximum of $V(\phi)$ through which the tunneling should occur. Thus, it does not seem possible to realize  open inflation scenario in the theories of the type of  (\ref{o5}).

As we have pointed out, this problem is very general. During several years which passed after the proposal of the one-field open inflation scenario \cite{BGT} nobody  proposed a single model where this scenario could be realized. In order to develop a successful one-field open universe scenario  one should solve several problems simultaneously. It is necessary to invent a potential which has a rather peculiar shape. It is not enough to make the potential very curved at its local minimum. The potential must be very curved also at the barrier and, more importantly, at the point to which the tunneling occurs. This seemed to imply that the effective potential should look like a cliff  with a sharp local minimum at the top, from which the scalar field  tunnels and falls down to a plateau. It is very difficult to invent such potentials. 
Moreover, in such models inflation may not begin immediately after the tunneling. One would expect that inflation  would begin  much later, after a period of rapid rolling of the scalar field, and perhaps even after a subsequent period of its oscillations. This would be quite different from the scenario anticipated in \cite{BGT}.  

We made several attempts to improve the situation in the model   (\ref{o5}) by adding  exponentially growing terms of the type of $\exp {C \phi^2}$ or $\exp {C \phi}$ with $C = O(1)$, which often appear in supergravity and string theory. We also tried to use scalar fields nonminimally coupled to gravity. All these attempts so far did not lead to a successful one-field open universe scenario. That is why it is very encouraging that one can overcome all problems discussed above in the context of a model with a very simple potential which we are going to describe. This potential   satisfies all required properties, and nevertheless inflation in this model  begins almost immediately after the tunneling, without any intermediate period of oscillations of the field $\phi$.

\section{A toy model of open inflation}
Let us consider a model with the effective potential of the following type:
\begin{equation}\label{toy}
V(\phi) =    {m^2\phi^2\over 2} \left (1+{\alpha^2\over  \beta^2+(\phi -v)^2}\right)
\end{equation}
Here $\alpha$ $\beta$ and $v$ are some constants; we will assume that $\beta \ll v$. The first term in this equation is the potential of the simplest chaotic inflation model ${m^2\phi^2\over 2}$.  The second term represents a peak of width $\beta$ with a maximum near $\phi = v$. The relative hight of  this peak with respect to the potential ${m^2\phi^2\over 2}$ is determined by the ratio ${\alpha^2\over \beta^2}$.  

\begin{figure}[Fig1]
 \hskip 1.5cm
\leavevmode\epsfysize=5cm \epsfbox{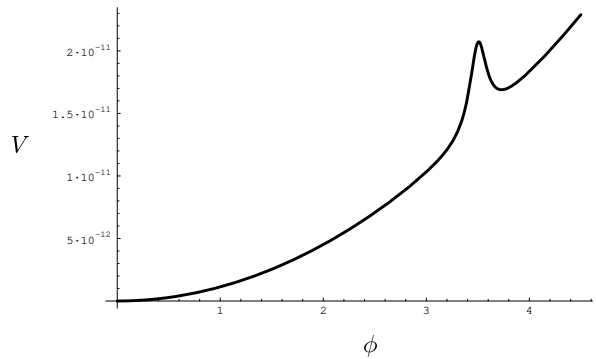}

\

\caption[Fig1]{\label{Pot} Effective potential in our toy model. All values are given in units where $M_p = 1$.}

\end{figure}

As an example, we will consider the theory with $m = 1.5 \times 10^{-6}$, which is necessary to have a proper amplitude of density perturbations during inflation in our model. We will take $v = 3.5$, which, as we will see,  will provide about 70 e-folds of inflation after the tunneling. By changing this parameter by few percent one can get any value of $\Omega$ from $0$ to $1$.  For definiteness, we will  take $ \beta^2 = 2 \alpha^2$, $\beta = 0.1$. This is certainly not a unique choice; other values of these parameters  can also lead to a successful open inflation scenario. The shape of the effective potential in this model is shown in Fig. \ref{Pot}.

\begin{figure}[Fig1]
 \hskip 1.5cm
\leavevmode\epsfysize=10cm \epsfbox{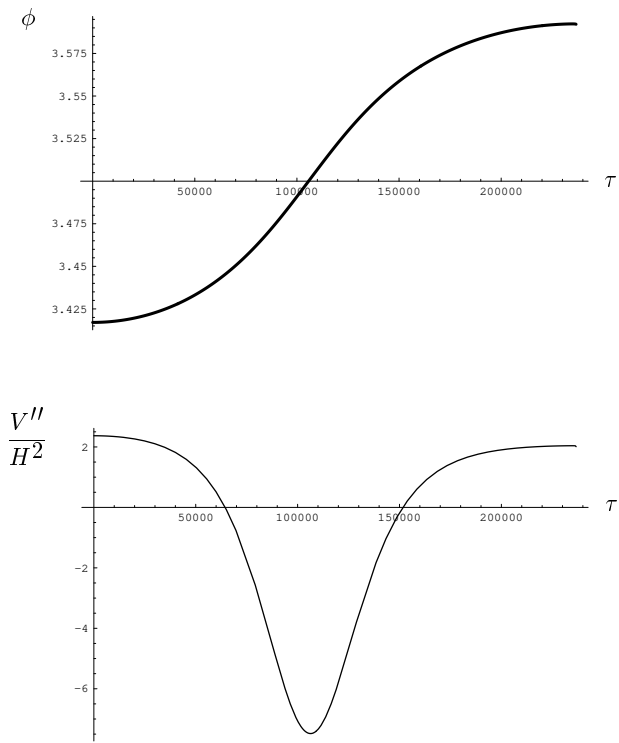}

\

\caption[Fig1]{\label{CDL} Coleman-De Luccia instanton in our model. The first panel shows the function $\phi(\tau)$, the second one demonstrates that, as we expected,  most of the time    during the tunneling one has $|V''| \gg H^2$.}

\end{figure}

As we see, this potential  coincides with ${m^2\phi^2\over 2}$ everywhere except a small vicinity of the point $\phi = 3.5$, but one cannot roll from $\phi > 3.5$ to $\phi < 3.5$ without tunneling through a sharp barrier. We have solved Eq. (\ref{eq1})  for this model numerically and  found that the Coleman-De Luccia instanton in this model does exist. It is shown in Fig. \ref{CDL}.

The upper panel of Fig. \ref{CDL} shows the CDL instanton $\phi(\tau)$.  Tunneling occurs from $\phi_{i} \approx 3.6$ to $\phi_{f} \approx 3.4$. The lower panel of Fig. \ref{CDL} shows the ratio $V''/H^2$. Almost everywhere along the trajectory $\phi(\tau)$ one has $|V''| > H^2$. That is exactly what we have expected on basis of our general arguments concerning CDL instantons. The condition $|V''| > H^2$ would be satisfied even better if we would take smaller value of the parameter $\alpha$.

Now we must check what happens after the tunneling. In order to do so, one should make an analytical continuation to Lorentzian space and study the time evolution of the scalar field $\phi(t)$ and of the scale factor $a(t)$.

Equations of motion for $\phi(t) $ and $a(t)$  in an open universe are
\begin{equation}\label{eq11}
\ddot\phi +3{\dot a\over a}\dot\phi=-V_{,\phi} \ , ~~~ 
 \ddot a= -{8\pi \over 3} a (\dot\phi^2 - V) \ .
\end{equation}
One should solve these equations with  boundary conditions $\dot\phi(0) = 0$, $a(0) = 0$, and $\dot a(0) = 1$.

\begin{figure}[Fig1]
 \hskip 1.5cm
\leavevmode\epsfysize=10cm \epsfbox{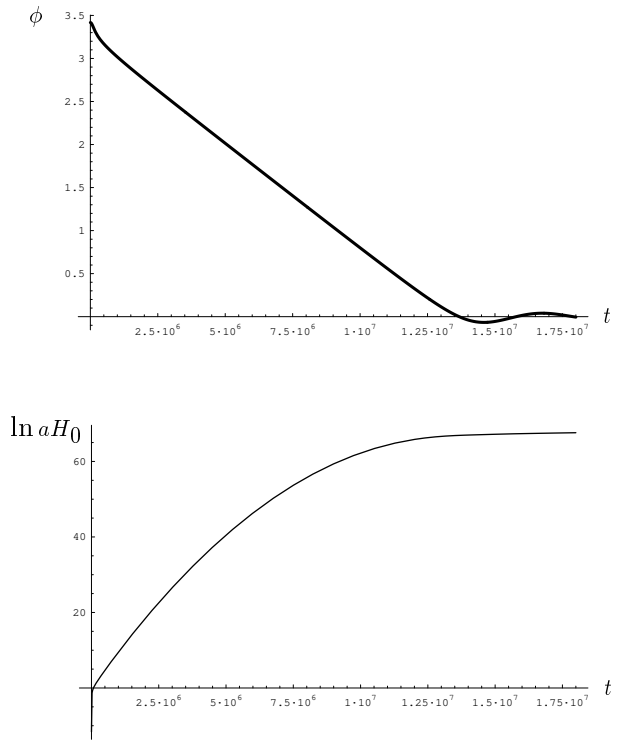}

\

\caption[Fig1]{\label{Inflation} Inflation after the tunneling. The upper panel shows how the field $\phi$ rolls down after the tunneling. The lower panel shows the growth of the logarithm of the scale factor. Here $H_0 =   \sqrt {8\pi V(\phi_f)/ 3}$.}

\end{figure}

Solutions of these equations for our model are shown in Fig. \ref{Inflation}. As we see, the scalar field slowly rolls down and then   oscillates near the minimum of the effective potential at $\phi = 0$. During the stage of the slow rolling, the scale factor expands approximately $e^{70}$ times. This is in a good agreement with the expression $\exp (2\pi\phi_f^2) \sim e^{70}$  for the degree of inflation in the theory with the effective potential ${m^2 \phi^2\over 2}$, with inflation beginning at $\phi_f \sim 3.4$ \cite{book}.
We have verified numerically that the universe would inflate   $e^{60}$ times if we would take $v = 3.2$.  Duration of inflation slightly changes also when we change other parameters of our model, because it affects the process of tunneling. This means that by a slight change of $v$ and other parameters one can  get any value of $\Omega$ in the interval $0 < \Omega < 1$.

Thus, in this model we do not have any problems associated with rapid rolling of the scalar field $\phi$ immediately after tunneling to the region with $V'' > H^2$. There are two reasons why this problem does not appear here. First of all, within the initial period of time $\sim H^{-1}$ after the bubble formation, the friction coefficient $3{\dot a\over a}$ in Eq. (\ref{eq11}) is especially large, because the scale factor of an open inflationary universe grows as $H^{-1} \sinh{Ht}$ rather than as $H^{-1} \exp Ht$. Therefore at this stage the speed of the scalar field $\phi$ grows more slowly that in flat space. Secondly, in our model the range of values of the field $\phi < \phi_f$ where $V'' > H^2$ is very narrow, so the field does not  acquire   speed which would be   large enough to destroy inflation during the short time when it rolls   from $\phi_f \sim 3.4$. This means that all problems outlined in the previous section can be successfully solved in our model.

Note, however, that the absence of a prolonged noninflationary stage after the tunneling is not a generic property of all one-field inflationary models, but a specific feature of our model. In different models, or in our model with a different set of parameters (e.g. with a much greater value of $\alpha/\beta$), one may have $V''(\phi_f) \gg H^2$ and  a long noninflationary stage after the tunneling. During this stage, the usual inflationary density perturbations will not be produced. 

Moreover, even in the version of our model considered above, $V''(\phi)$ remains greater than $H^2$ during the first 3 e-folds of inflation. This means that the usual inflationary perturbations are not produced inside the bubble during this time. Also, the value of $|\dot\phi|$ during the first few e-folds of inflation  is approximately twice as large as its value in the beginning of the asymptotic inflationary regime where the field $\phi$ decreases linearly, see Fig. \ref{Inflation}. Since usually density perturbations are inversely proportional to $|\dot\phi|$, this effect  may  lead to an additional     suppression of density perturbations on the scale of the horizon, which may  affect correspondingly the magnitude of the microwave background anisotropy in open inflation. 

\begin{figure}[Fig1]
 \hskip 1.5cm
\leavevmode\epsfysize=4.8cm \epsfbox{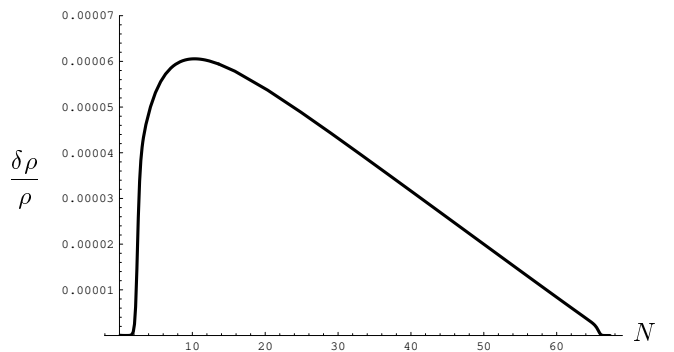}

\

\caption[Fig1]{\label{Pert} Density perturbations ${\delta\rho\over \rho}$  produced inside the bubble $N$\, e-folds after the open universe creation. If $N = O(1)$ corresponds to density perturbations on the horizon scale $\sim 10^{28}$ cm, then the maximum of the spectrum appears on scale which is about four orders of magnitude smaller. However, these density perturbations should be supplemented by the bubble wall perturbations and supercurvature fluctuations, which may alter the final result for ${\delta\rho\over \rho}$ at small $N$.}

\end{figure}

Investigation of density perturbations in this scenario is rather complicated because it involves several different contributions.   We will not perform a complete investigation of this question here but rather make an estimate of the standard contribution of quantum fluctuations produced inside the bubble. To evaluate it,   we plotted  the function ${\delta\rho\over \rho}   = {6\over 5} {H^2(\phi)\over 2\pi |\dot\phi|}$, which would correspond to density perturbations    in a flat cold dark matter dominated universe in normalization of  Ref. \cite{book}.  In an open inflationary   universe this should be somewhat corrected during the very first stages  of inflation, at $N = O(1)$ \cite{YST,Bellido}. However, as we already pointed out, during the first three e-folds inflationary fluctuations of the field $\phi$ are not produced inside the bubble. Meanwhile one may expect that the standard flat-space expression gives correct results at $N > 3$,   when the scale factor  $H^{-1} \sinh{Ht}$ approaches the flat space regime $H^{-1} \exp Ht$.  Fig.~\ref{Pert} shows the magnitude of perturbations produced at the moment corresponding to   $N$   e-folds of inflation after the open universe formation. As we see,  ${\delta\rho\over \rho} $ has a deep minimum  at $N \lesssim 4$, which appears because of the large curvature of $V(\phi)$ and large speed of the field $\phi$ soon after the tunneling.  Then the  magnitude of density perturbations approaches its maximum  at $N \sim 10$, and after that its behavior becomes the same as in the flat space chaotic inflation in the theory ${m^2\over 2}\phi^2$.   If one changes parameters  and takes, for example, $v = 3.85$, and increases $\alpha^2$ ten times, then the total duration of inflation does not change, and the spectrum of perturbations looks very similar to the one shown in Fig.~\ref{Pert}, but its maximum appears at somewhat greater values of $N$  (at $N \sim 15$), which corresponds to a smaller length scale.  Roughly speaking, if one interprets perturbations produced immediately after the creation of the open universe   (at $N = O(1)$) as  perturbations on the horizon scale $l \sim 10^{28}$ cm, then the maximum  at $N \sim 10$ would correspond to scale $\sim 10^{24}$ cm, and the maximum  at $N \sim 15$ would correspond to  scale $\sim 10^{22}$ cm, which is similar to galaxy scale.

The mechanism of suppression of  large scale  density perturbations  described above is  a generic feature of the one-field   open inflation models based on   tunneling and bubble formation. However, one should check whether this suppression can be compensated by     other sources of perturbations. Indeed, in  addition to the standard  scalar field perturbations produced inside the bubble one should also consider   long wavelength perturbations produced outside the bubble (supercurvature modes). When the bubble wall grows, these perturbations penetrate  the bubble. Naively, one would not expect any supercurvature fluctuations in our model because of the condition $|V''| > H^2$. However, these perturbations do exist, at least for our choice of the parameters. Indeed, one can check that $|V''| \ll  H^2$ at the local minimum of the effective potential, even though the condition $|V''| > H^2$ is satisfied during the tunneling and soon after it. These two statements are  consistent with each other because the CDL instanton describes tunneling not exactly from the false vacuum, where $|V''| \ll  H^2$, but slightly away from it, where one has $|V''| >  H^2$ \cite{min}.

Fluctuations of the 
field $\phi$   outside the bubble are proportional to the large Hubble constant prior to the tunneling, and therefore they have a slightly  greater amplitude than those   inside the bubble. 
In addition to  scalar  perturbations, one may also consider tensor perturbations associated with perturbations of the bubble wall. Preliminary estimates of the amplitude of these perturbations   based on the methods  of Refs. \cite{Garriga,Sasaki,Bellido} indicate that the magnitude of these perturbations in our model can be consistent with observational data. 

The contribution of  supercurvature   modes and tensor perturbations may compensate the suppression of the usual density perturbations in the long-wavelength limit, i.e. at  $N \sim 1$. However, unless the amplitude of supercurvature    and tensor perturbations is extremely large, one may expect that the total expression for density perturbations will still have a maximum at large $N$ even if all contributions to density perturbations are taken into account. This possibility deserves further investigation because the existence of a maximum in the spectrum of density perturbations at an intermediate scale could serve as a distinctive signature of the one-field open inflation models described above.

\section{Discussion}
 Recent developments in observational cosmology suggest that one may not need to invent complicated versions of inflationary theory describing the universe with $\Omega < 1$. However, since we still do not know for sure the true value of $\Omega$ (including the vacuum energy contribution $\Omega_\Lambda$), it would be better to be on a safe side and to have  some  inflationary models where  $\Omega$ can be smaller than $1$. Also, independently of any practical purposes, the possibility that an infinite open universe can be created by tunneling within a finite region of space is one of the most beautiful effects of general relativity. It would be a pity if it were impossible or extremely complicated to realize it. 

The toy model proposed in this paper provides a  simple realization of the one-field open universe scenario. It resolves all problems outlined in Sect. \ref{problems} and 
demonstrates that such models can be  viable. In a certain sense, it provides the  last missing link in the existence proof for open inflation. It still remains to be seen whether our model can pass all cosmological tests. In particular, one should perform a detailed study of density perturbations and   CMB anisotropy produced in this model.  To make one-field open inflation models  completely realistic, one would need to find a physical mechanism for appearance of a sharp peak in the effective potential. One of the  mechanisms which may lead to a peak in $V(\phi)$  is   an emergence of a strong coupling regime in the Yang-Mills sector of the theory  when $\phi$ approaches $v$. Then the energy density acquires additional contribution   from the condensates of the type of $\langle F^a_{\mu \nu} F_a^{\mu \nu}\rangle$ and  $\langle \bar \psi\psi\rangle$, which vanishes away from the strong coupling region.  Perhaps it will be easier to identify a possible origin of the peak in $V(\phi)$ in the context of hybrid inflation \cite{Hybrid}. But we do not want to speculate about it now. The main goal of this paper was very limited. We   wanted to show that  viable models of  open inflation do exist, and that despite our earlier expectations, the effective potentials in such models may look very simple, see Fig. \ref{Pot}.  It is certainly much easier to construct inflationary models with $\Omega = 1$, which remains one of the main predictions of most inflationary models.  But in the absence of any non-inflationary theory which would explain homogeneity of an open universe, it is good to know that the general idea of inflation is robust enough to incorporate models with any possible value of $\Omega$.

\subsection*{Acknowledgments}

It is a pleasure to thank  R. Bousso, M. Bucher, G.~Dvali, J. Garc\'{\i}a--Bellido, J. Garriga, N. Kaloper, L.~Kofman and M. Sasaki   for useful
discussions. I am also grateful to ``Poble Espanyol,"  Barcelona, for the warm hospitality during the time when the main part of this work was completed. This work   was supported in part by NSF grant    PHY-9870115.

\end{document}